\documentstyle[twoside,psfig]{article}

\catcode`\@=11
\long\def\@makefntext#1{
\protect\noindent \hbox to 3.2pt {\hskip-.9pt  
$^{{\eightrm\@thefnmark}}$\hfil}#1\hfill}		

\def\thefootnote{\fnsymbol{footnote}}
\def\@makefnmark{\hbox to 0pt{$^{\@thefnmark}$\hss}}	
	
\def\ps@myheadings{\let\@mkboth\@gobbletwo
\def\@oddhead{\hbox{}
\rightmark\hfil\eightrm\thepage}   
\def\@oddfoot{}\def\@evenhead{\eightrm\thepage\hfil
\leftmark\hbox{}}\def\@evenfoot{}
\def\sectionmark##1{}\def\subsectionmark##1{}}

\oddsidemargin=\evensidemargin
\renewcommand{\thefootnote}{\fnsymbol{footnote}}

\newcounter{sectionc}\newcounter{subsectionc}\newcounter{subsubsectionc}
\renewcommand{\section}[1] {\vspace{12pt}\addtocounter{sectionc}{1} 
\setcounter{subsectionc}{0}\setcounter{subsubsectionc}{0}\noindent 
	{\tenbf\thesectionc. #1}\par\vspace{5pt}}
\renewcommand{\subsection}[1] {\vspace{12pt}\addtocounter{subsectionc}{1} 
	\setcounter{subsubsectionc}{0}\noindent 
	{\bf\thesectionc.\thesubsectionc. {\kern1pt \bfit #1}}\par\vspace{5pt}}
\renewcommand{\subsubsection}[1] {\vspace{12pt}\addtocounter{subsubsectionc}{1}
	\noindent{\tenrm\thesectionc.\thesubsectionc.\thesubsubsectionc.
	{\kern1pt \tenit #1}}\par\vspace{5pt}}
\newcommand{\nonumsection}[1] {\vspace{12pt}\noindent{\tenbf #1}
	\par\vspace{5pt}}

\newcounter{appendixc}
\newcounter{subappendixc}[appendixc]
\newcounter{subsubappendixc}[subappendixc]
\renewcommand{\thesubappendixc}{\Alph{appendixc}.\arabic{subappendixc}}
\renewcommand{\thesubsubappendixc}
	{\Alph{appendixc}.\arabic{subappendixc}.\arabic{subsubappendixc}}

\renewcommand{\appendix}[1] {\vspace{12pt}
        \refstepcounter{appendixc}
        \setcounter{figure}{0}
        \setcounter{table}{0}
        \setcounter{lemma}{0}
        \setcounter{theorem}{0}
        \setcounter{corollary}{0}
        \setcounter{definition}{0}
        \setcounter{equation}{0}
        \renewcommand{\thefigure}{\Alph{appendixc}.\arabic{figure}}
        \renewcommand{\thetable}{\Alph{appendixc}.\arabic{table}}
        \renewcommand{\theappendixc}{\Alph{appendixc}}
        \renewcommand{\thelemma}{\Alph{appendixc}.\arabic{lemma}}
        \renewcommand{\thetheorem}{\Alph{appendixc}.\arabic{theorem}}
        \renewcommand{\thedefinition}{\Alph{appendixc}.\arabic{definition}}
        \renewcommand{\thecorollary}{\Alph{appendixc}.\arabic{corollary}}
        \renewcommand{\theequation}{\Alph{appendixc}.\arabic{equation}}
        \noindent{\tenbf Appendix \theappendixc #1}\par\vspace{5pt}}
\newcommand{\subappendix}[1] {\vspace{12pt}
        \refstepcounter{subappendixc}
        \noindent{\bf Appendix \thesubappendixc. {\kern1pt \bfit #1}}
	\par\vspace{5pt}}
\newcommand{\subsubappendix}[1] {\vspace{12pt}
        \refstepcounter{subsubappendixc}
        \noindent{\rm Appendix \thesubsubappendixc. {\kern1pt \tenit #1}}
	\par\vspace{5pt}}

\newcommand{\textlineskip}{\baselineskip=13pt}
\newcommand{\smalllineskip}{\baselineskip=10pt}

\def\eightcirc{
\begin{picture}(0,0)
\put(4.4,1.8){\circle{6.5}}
\end{picture}}
\def\eightcopyright{\eightcirc\kern2.7pt\hbox{\eightrm c}} 

\newcommand{\copyrightheading}[1]
	{\vspace*{-2.5cm}\smalllineskip{\flushleft
	}}

\newcommand{\publisher}[2]{{\begin{center}\footnotesize\smalllineskip 
	\end{center}
	}}

\def\abstracts#1#2#3{{
	\centering{\begin{minipage}{4.5in}\footnotesize\baselineskip=10pt
	\parindent=0pt #1\par 
	\parindent=15pt #2\par
	\parindent=15pt #3
	\end{minipage}}\par}} 

\newcommand{\bibit}{\nineit}
\newcommand{\bibbf}{\ninebf}
\renewenvironment{thebibliography}[1]
	{\frenchspacing
	 \ninerm\baselineskip=11pt
	 \begin{list}{\arabic{enumi}.}
        {\usecounter{enumi}\setlength{\parsep}{0pt}     
	 \setlength{\leftmargin 12.7pt}{\rightmargin 0pt} 
         \setlength{\itemsep}{0pt} \settowidth
	{\labelwidth}{#1.}\sloppy}}{\end{list}}

\newcounter{itemlistc}
\newcounter{romanlistc}
\newcounter{alphlistc}
\newcounter{arabiclistc}

\newcommand{\fcaption}[1]{
        \refstepcounter{figure}
        \setbox\@tempboxa = \hbox{\footnotesize Fig.~\thefigure. #1}
        \ifdim \wd\@tempboxa > 5in
           {\begin{center}
        \parbox{5in}{\footnotesize\smalllineskip Fig.~\thefigure. #1}
            \end{center}}
        \else
             {\begin{center}
             {\footnotesize Fig.~\thefigure. #1}
              \end{center}}
        \fi}

\newcommand{\tcaption}[1]{
        \refstepcounter{table}
        \setbox\@tempboxa = \hbox{\footnotesize Table~\thetable. #1}
        \ifdim \wd\@tempboxa > 5in
           {\begin{center}
        \parbox{5in}{\footnotesize\smalllineskip Table~\thetable. #1}
            \end{center}}
        \else
             {\begin{center}
             {\footnotesize Table~\thetable. #1}
              \end{center}}
        \fi}

\def\@citex[#1]#2{\if@filesw\immediate\write\@auxout
	{\string\citation{#2}}\fi
\def\@citea{}\@cite{\@for\@citeb:=#2\do
	{\@citea\def\@citea{,}\@ifundefined
	{b@\@citeb}{{\bf ?}\@warning
	{Citation `\@citeb' on page \thepage \space undefined}}
	{\csname b@\@citeb\endcsname}}}{#1}}

\newif\if@cghi
\def\cite{\@cghitrue\@ifnextchar [{\@tempswatrue
	\@citex}{\@tempswafalse\@citex[]}}
\def\citelow{\@cghifalse\@ifnextchar [{\@tempswatrue
	\@citex}{\@tempswafalse\@citex[]}}
\def\@cite#1#2{{$\null^{#1}$\if@tempswa\typeout
	{IJCGA warning: optional citation argument 
	ignored: `#2'} \fi}}

\def\pmb#1{\setbox0=\hbox{#1}
	\kern-.025em\copy0\kern-\wd0
	\kern.05em\copy0\kern-\wd0
	\kern-.025em\raise.0433em\box0}

\def\fnt#1#2{\footnotetext{\kern-.3em
	{$^{\mbox{\scriptsize #1}}$}{#2}}}

\def\fpage#1{\begingroup
\voffset=.3in
\thispagestyle{empty}\begin{table}[b]\centerline{\footnotesize #1}
	\end{table}\endgroup}

\def\runninghead#1#2{\pagestyle{myheadings}
\markboth{{\protect\footnotesize\it{\quad #1}}\hfill}
{\hfill{\protect\footnotesize\it{#2\quad}}}}
\font\tenrm=cmr10
\font\tenit=cmti10 
\font\tenbf=cmbx10
\font\bfit=cmbxti10 at 10pt
\font\ninerm=cmr9
\font\nineit=cmti9
\font\ninebf=cmbx9
\font\eightrm=cmr8

\def\qed{\hbox{${\vcenter{\vbox{			
   \hrule height 0.4pt\hbox{\vrule width 0.4pt height 6pt
   \kern5pt\vrule width 0.4pt}\hrule height 0.4pt}}}$}}

\renewcommand{\thefootnote}{\fnsymbol{footnote}}	

\begin{document}
\setlength{\textheight}{7.7truein}  

\runninghead{M. Freund \& T. Ohlsson}{Matter
Enhanced Neutrino Oscillations $\ldots$}

\normalsize\textlineskip
\thispagestyle{empty}
\setcounter{page}{1}

\copyrightheading{}			

\vspace*{0.88truein}

\fpage{1}
\centerline{\bf MATTER ENHANCED NEUTRINO OSCILLATIONS}
\baselineskip=13pt
\centerline{\bf WITH A REALISTIC EARTH DENSITY PROFILE}
\vspace*{0.37truein}
\centerline{\footnotesize MARTIN FREUND\footnote{E-mail: {\tt
martin.freund@physik.tu-muenchen.de}}}
\baselineskip=12pt
\centerline{\footnotesize\it Institut f{\"u}r Theoretische Physik,
Physik-Department, Technische Universit{\"a}t M{\"u}nchen,
James-Franck-Stra\ss{}e}
\baselineskip=10pt
\centerline{\footnotesize\it DE-85748 Garching, Germany}
\vspace*{10pt}

\centerline{\footnotesize TOMMY OHLSSON\footnote{E-mail: {\tt
tommy@theophys.kth.se}}}
\baselineskip=12pt
\centerline{\footnotesize\it Division of Mathematical Physics,
Theoretical Physics, Department of Physics, Royal Institute of Technology}
\baselineskip=10pt
\centerline{\footnotesize\it SE-100 44 Stockholm, Sweden}
\vspace*{0.225truein}

\vspace{1.05cm}
\publisher{(received date)}{(revised date)}

\vspace*{0.21truein}
\abstracts{We have investigated matter enhanced neutrino oscillations with a
mantle-core-mantle step function and a realistic Earth matter density profile
in both a two and a three neutrino scenario. We found that the realistic
Earth matter density profile can be well approximated with the
mantle-core-mantle step function and that there could be an influence
on the oscillation channel $\nu_\mu \to \nu_\tau$ due to resonant
enhancement of one of the mixing angles.}{}{}



\setcounter{footnote}{0}
\renewcommand{\thefootnote}{\alph{footnote}}

\vspace*{1pt}\textlineskip	
\section{Introduction}	
\label{sec:intro}
\vspace*{-0.5pt}
\noindent
Step function approximations of the Earth matter density profile have
been used by several authors, see {\it e.g.}
Refs. \cite{nico88,giun98,liu98,petc98}, for the analysis of matter effects
in neutrino oscillation experiments. In order to investigate the
quality of this approximation, we have developed methods to
numerically calculate the transition probability amplitudes in a
realistic matter density profile. Calculations were performed in a two
and a three neutrino framework. The neutrino traveling path length $L$
was divided up into several small intervals. In each of these
intervals, in which we assumed constant matter density, we computed
the effective mixing parameters. Combining the probability amplitudes
of each interval leads to the total probability amplitude from which
we obtained the transition probabilities.

\begin{figure}[htbp] 
\vspace*{13pt}
\centerline{\psfig{file=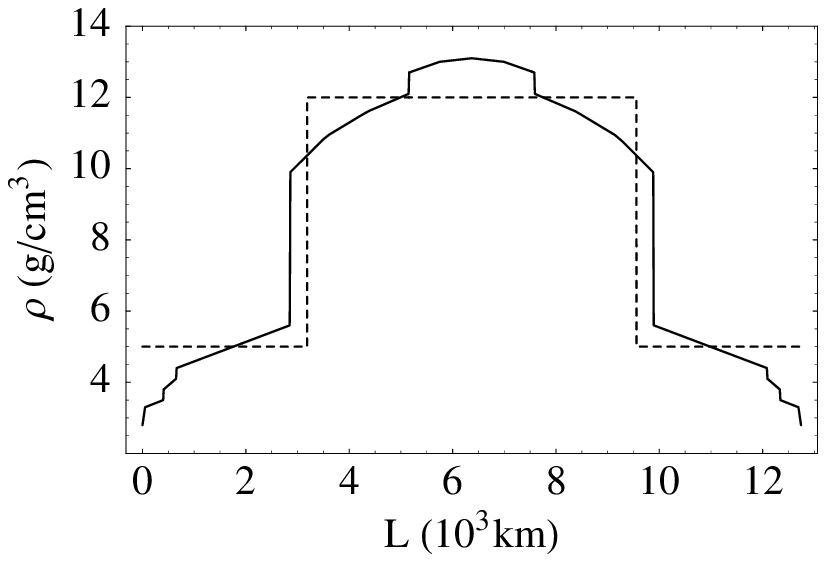,height=5cm}} 
\vspace*{13pt}
\fcaption{The matter density profiles. Realistic matter density profile (solid
curve) and mantle-core-mantle step function (dashed curve).}
\label{fig:1}
\end{figure}

We have used three different matter density profiles, two model-like and one
realistic: a constant profile just to show the resonance behavior, a
step function describing the mantle-core-mantle structure of the
Earth, and a realistic Earth density profile \cite{stac77}. 
See Fig.~\ref{fig:1} for the different density profiles. For the
mantle-core-mantle density profile, we chose $\rho = 5 \; {\rm
g/cm^3}$ for the mantle, $\rho = 12 \; {\rm g/cm^3}$ for the
core, and a core-width equal to half of the Earth's diameter.  

In our study we used the mass squared difference as suggested
by atmospheric neutrino data from the Super-Kamiokande Collaboration
obtained within a two neutrino analysis \cite{scho99}
$$
\Delta m^2 \simeq 3.2 \cdot 10^{-3} \; {\rm eV^2}.
$$

For the constant density profiles in the case of two neutrino
oscillations, we considered the well-known Mikheyev--Smirnov--Wolfenstein
(MSW) \cite{mikh85} resonance condition
\begin{equation}
\cos 2\theta - \frac{A}{\Delta m^2} = 0, \quad A =
A(\rho,E_\nu) = \frac{2 \sqrt{2} G_F Y}{m_N} \rho E_\nu,
\label{eq:MSW}
\end{equation}
where $\theta$ and $\Delta m^2$ are the usual mixing parameters for two
neutrino oscillations, $G_F$ is the Fermi weak coupling constant, $Y$
($\simeq 1/2$) is the average number of electrons per nucleon, $m_N$ is
the nucleon mass, $\rho$ is the matter density, and $E_\nu$ is the
neutrino energy. Using Eq.~(\ref{eq:MSW}) when $\theta$ approaches
zero, we obtained the resonant energies $E_\nu \simeq 3.6 \; {\rm
GeV}$ for the core and $E_\nu \simeq 8.5 \; {\rm GeV}$ for the mantle.

\vspace*{1pt}\textlineskip	
\section{Two Neutrino Calculation}	
\label{sec:two}
\begin{figure}[htbp] 
\vspace*{13pt}
\centerline{\psfig{file=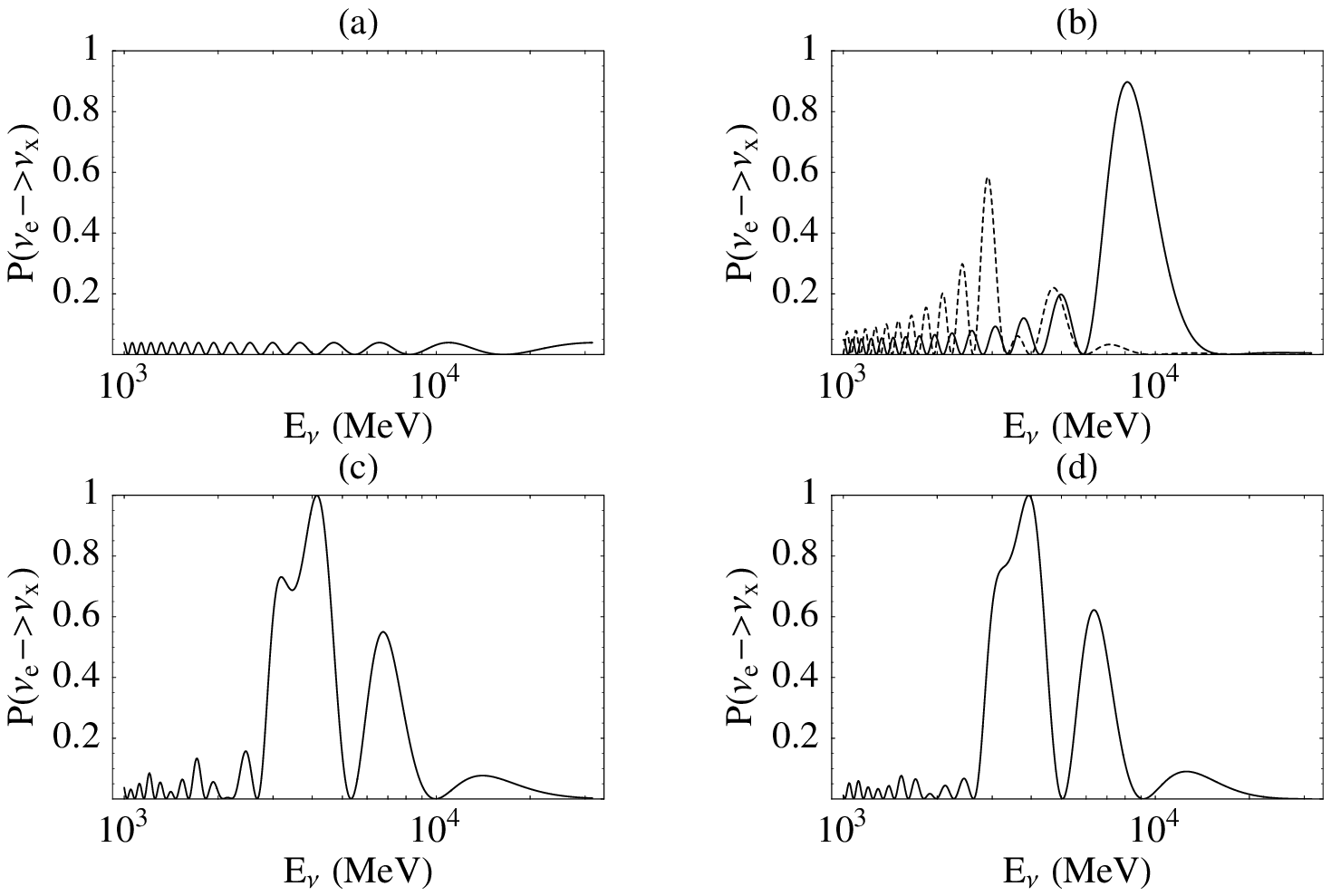,height=5cm}} 
\vspace*{13pt}
\fcaption{The transition probability $P(\nu_e \to \nu_x)$ as a function
of the neutrino energy $E_\nu$. (a) zero matter
density, (b) constant matter densities (mantle [solid curve] and core
[dashed curve]), (c) mantle-core-mantle step function, and (d) 
realistic Earth matter density profile.}
\label{fig:2}
\end{figure}

\vspace*{-0.5pt}
\noindent
As mentioned above, we used $\Delta m^2 = 3.2 \cdot 10^{-3} \; {\rm
eV^2}$ for the mass squared difference. Being interested in
oscillation enhancement, we chose a small mixing angle $\theta =
0.1$. Using these mixing parameters, we calculated the transition
probability $P(\nu_e \to \nu_x)$, where $x$ could be $\mu$ or $\tau$,
for neutrinos going through the whole Earth (nadir angle zero). The
results for the different matter density profiles discussed above are
displayed in Fig.~\ref{fig:2} as a function of the neutrino energy $E_\nu$.

In the absence of matter (Fig.~\ref{fig:2}~(a)), there is a constant
maximal probability of $\sin^2 2\theta$ without any enhancement. For
constant matter densities (Fig.~\ref{fig:2}~(b)), mantle and core, we obtained
oscillation enhancements at the expected neutrino energies of $E_\nu
\simeq 8.4 \; {\rm GeV}$ and $E_\nu \simeq 3.5 \; {\rm GeV}$, respectively. 
Figure~\ref{fig:2}~(c) shows the spectrum for the mantle-core-mantle
model. The resonance spectrum is interpreted in the following way
\cite{liu98,petc98,ermi86}: The main resonance is due to interference
and is therefore not MSW-like. The shoulder to the left on top of the main
peak is caused by the core and the peak to the right is caused by the
mantle. Both these resonances are typically MSW-like.
As seen from Figs.~\ref{fig:2}~(c)-(d), there is no large qualitative
difference between the matter enhancements obtained using the step function and
the realistic Earth density profiles. It turns out that the results for the
step function profile strongly depend on the widths of the mantle and
the core. 

The two neutrino model is not suitable for studying matter effects on
the $\nu_\mu \to \nu_\tau$ channel. To obtain more general results, we
performed this investigation also in a three neutrino framework.

\vspace*{1pt}\textlineskip	
\section{Three Neutrino Calculation}	
\label{sec:three}
\vspace*{-0.5pt}
\noindent
We used a hierarchal mass scheme $\Delta m^2_{21} \ll \Delta m^2_{32}
\simeq \Delta m^2_{31}$, where the large mass squared
difference should describe atmospheric oscillations and the small mass
squared difference should be responsible for solar oscillations.
As in the two neutrino analysis, we again chose the large mass squared
difference as $\Delta m_{32}^2 = 3.2 \cdot 10^{-3} \; {\rm eV^2}$. The
small mass squared difference $\Delta m_{21}^2$ was set to
zero under the assumption that these transitions play no role at
length scales of the diameter of the Earth and smaller.

We also used the standard parameterization for the neutrino mixing
matrix with the mixing angles $\theta_{12}$, $\theta_{13}$, and
$\theta_{23}$. To accommodate the atmospheric neutrino result, we chose
$\theta_{23}=\pi/4$ giving the following vacuum transition probabilities
\begin{eqnarray}
P(\nu_e \to \nu_e) &=& 1 - \sin^2 2\theta_{13} \sin^2 \frac{\Delta
m_{32}^2 L}{4 E_\nu}, \label{eq:ee}\\
P(\nu_e \to \nu_\mu) &=& P(\nu_e \to \nu_\tau) = \frac{1}{2} \sin^2
2\theta_{13} \sin^2 \frac{\Delta m_{32}^2 L}{4 E_\nu}, \label{eq:em}\\
P(\nu_\mu \to \nu_\mu) &=& 1 - \cos^2
\theta_{13}\left(2-\cos^2\theta_{13}\right) \sin^2 
\frac{\Delta m_{32}^2 L}{4 E_\nu}, \label{eq:mm}\\
P(\nu_\mu \to \nu_\tau) &=& \cos^4 \theta_{13} \sin^2 \frac{\Delta
m_{32}^2 L}{4 E_\nu} \label{eq:mt}.
\end{eqnarray}

The parameters relevant for oscillations involving $\Delta m^2_{32}$
are $\theta_{13}$ and $\theta_{23}$. The magnitude of $\theta_{12}$
plays no crucial role even in presence of matter effects. For the
numerical calculations we set $\theta_{12} = 0$ but also a maximal
angle would not change our obtained results.

Reactor experiments, the solar neutrino deficit and the atmospheric
neutrino anomaly all indicate that the mixing angle $\theta_{13}$
should be small. The most stringent upper bound of $\sin^2
2\theta_{13} = 0.10$ (valid for 
$\Delta m^2 > 0.7 \cdot 10^{-3} \; {\rm eV^2}$) is coming from the
CHOOZ experiment \cite{apol98}. As in the two neutrino calculation we chose
$\theta_{13}=0.1$, which is obviously below the CHOOZ upper bound.  

\begin{figure}[htbp] 
\vspace*{13pt}
\centerline{\psfig{file=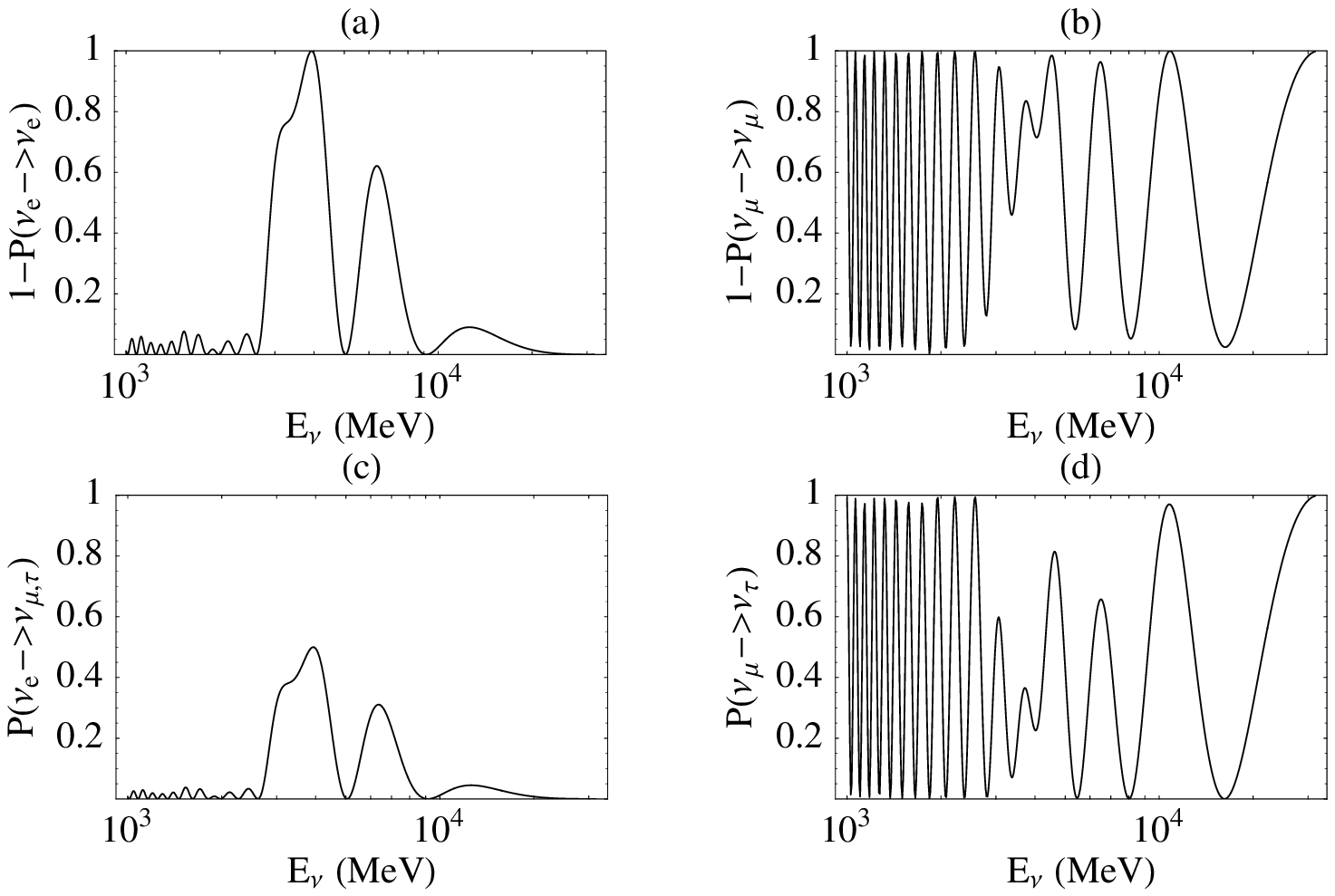,height=5cm}} 
\vspace*{13pt}
\fcaption{Different transition probabilities as functions of the
neutrino energy $E_\nu$ for the realistic Earth matter density
profile. (a) $1 - P(\nu_e \to \nu_e)$, (b) $1 - P(\nu_\mu \to
\nu_\mu)$, (c) $P(\nu_e \to \nu_{\mu,\tau})$, and (d) $P(\nu_\mu \to
\nu_\tau)$.}
\label{fig:3}
\end{figure}

The oscillation probability $P(\nu_e \to \nu_e)$ is given by an
effective two neutrino formula with $\theta_{13}$ being the relevant
mixing angle. Thus, the two neutrino model discussed above should give a good
approximation for the oscillation enhancement obtained in this
channel. The resonant energies in the three neutrino model are
expected to be approximately the same as calculated in the two neutrino model.
This can be seen in Fig.~\ref{fig:3}~(a), where the
oscillation probability $1-P(\nu_e \to \nu_e)$ is shown as a function
of the neutrino energy $E_\nu$ for the realistic Earth density profile.
Figure~\ref{fig:3}~(a) is very similar to the corresponding two
neutrino case, see Fig.~\ref{fig:2}~(d).

From Eq.~(\ref{eq:em}) it can be seen, that in the non-resonant region,
electron neutrino disappearance is caused in equal amounts by the
transitions $\nu_e \to \nu_\mu$ and $\nu_e \to \nu_\tau$. This remains true
also in the presence of matter as is shown in Fig.~\ref{fig:3}~(c). 

Due to the resonance enhancement of $\theta_{13}$, we expect a drop in
the amplitude for the oscillation probabilities $1 - P(\nu_\mu \to
\nu_\mu)$ and $P(\nu_\mu \to \nu_\tau)$, see Figs.~\ref{fig:3}~(b) and
(d), respectively. Hence, the oscillation channel  $\nu_\mu \to
\nu_\tau$ relevant for the atmospheric neutrino anomaly is influenced
indirectly by matter effects through the enhancement of
$\theta_{13}$.\footnote{In two neutrino analyses, the $\nu_\mu \to
\nu_\tau$ oscillation is usually expected not to be influenced by
matter effects as there is no coupling to matter via the charged
current interactions, which are present only for the electron flavor.} 

In similarity to the two neutrino scheme, we found no sizable differences
for the mantle-core-mantle model and the realistic Earth matter
density profile.

\begin{figure}[htbp] 
\vspace*{13pt}
\centerline{\psfig{file=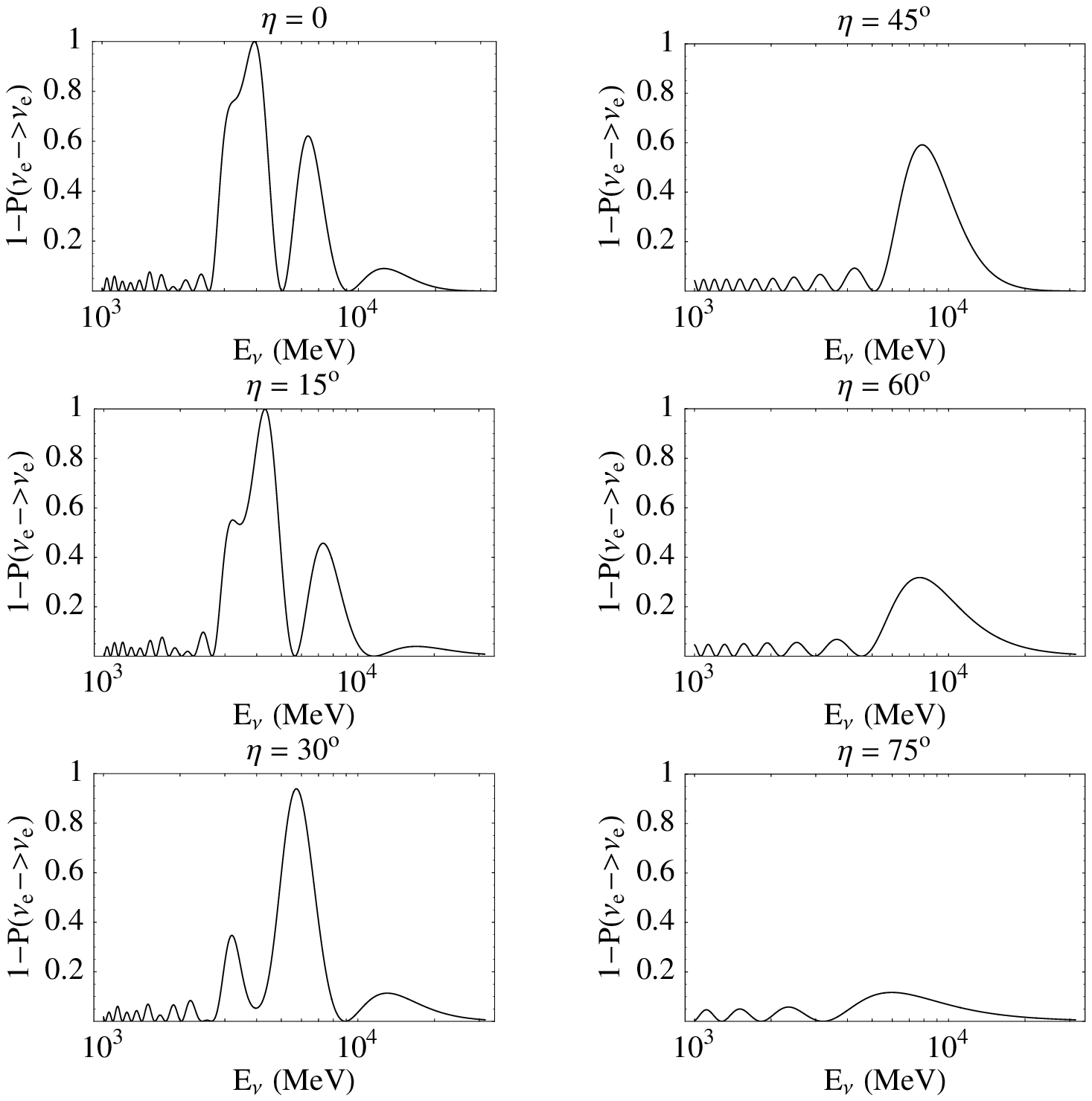,height=7.5cm}} 
\vspace*{13pt}
\fcaption{The probability $1-P(\nu_e \to \nu_e)$ as a function of the
neutrino energy $E_\nu$ for different nadir angles.}
\label{fig:4}
\end{figure}

We also investigated the nadir angle dependence in the three
neutrino case. The results are shown in Fig.~\ref{fig:4} for different
nadir angles $\eta$.
Note that for increasing nadir angle, the resonance for the core
shrinks and the mantle starts to dominate the resonance spectrum. For angles
exceeding $\eta \sim 34^\circ$ neutrinos do no longer traverse the
core. The resonance is then centered around the resonant energy $E_\nu \simeq
8.5 \; {\rm GeV}$ corresponding to the matter density of the mantle
(see Fig.~\ref{fig:4} for $\eta = 45^\circ$).
For larger nadir angles also the mantle resonance fades away due to the
fact that the overall traveling path length of the neutrinos in the mantle
decreases to zero.

To achieve a large and well defined effect of
matter enhancement, we thus propose to further study nadir angles just above
the threshold where the traveling length is maximal through the mantle
of the Earth.

\vspace*{1pt}\textlineskip	
\section{Atmospheric Neutrino Simulation}	
\label{sec:AT}
\vspace*{-0.5pt}
\noindent
To describe the quality of the step function approximation in a more 
intuitive way, we simulated a typical atmospheric neutrino experiment
(like the Super-Kamiokande experiment) using the Honda {\it et al.}
atmospheric neutrino fluxes \cite{hond95}. 
\begin{figure}[htbp] 
\vspace*{13pt}
\centerline{\psfig{file=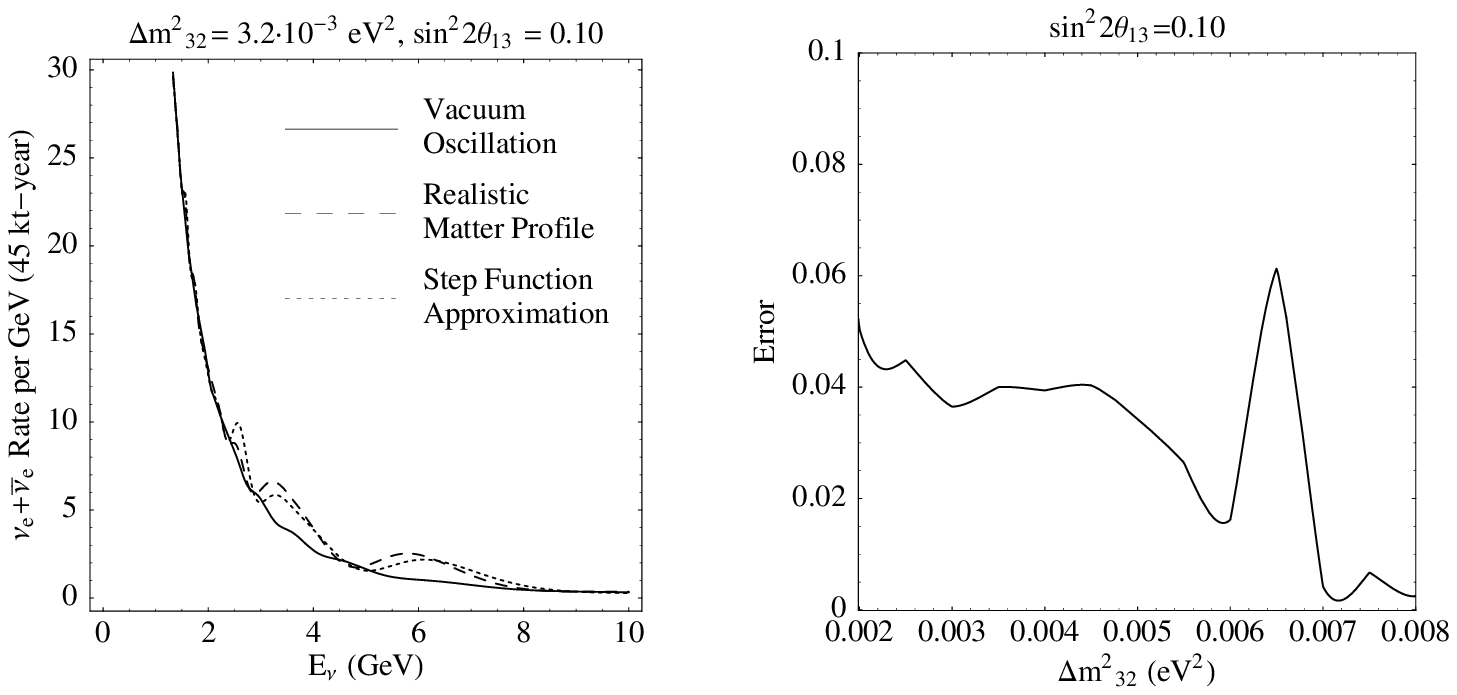,height=4.5cm}} 
\vspace*{13pt}
\fcaption{The left plot shows the energy spectrum of the electron
neutrino-like events in an atmospheric neutrino detector for vacuum
oscillation (solid), the realistic Earth matter density profile (dashed), and
the step function approximation (dotted). The right plot shows the
error of the step function approximation for different $\Delta m^2_{32}$.}
\label{fig:5}
\end{figure}
The result is shown in Fig.~\ref{fig:5}. The left plot
shows the energy spectrum of the core-crossing neutrino bin 
($\cos \eta = 0.8 \div 1.0$) in a $45\mathrm{kt}$-detector during one year 
of running. The figure, which is based on $\sin^2 2\theta_{23} = 1$,
$\sin^2 2\theta_{13} = 0.10$, and 
$\Delta m^2 = 3.2\cdot 10^{-3} \; {\rm eV}^2$, shows the cases: vacuum 
oscillation (solid), realistic matter profile (dashed), and step function 
approximation (dotted). The enhancements compared to the vacuum case
are due to resonant matter effects. The right plot tries to quantify
the difference of the two matter profiles. It shows the ``error''
of the step function approximation in the calculation of the
total number of electron neutrino-like events in the core-crossing bin. The
error is defined as the ratio $|(N_{\mathrm{RMP}} - N_{\mathrm{SFA}})/
(N_{\mathrm{RMP}} - N_{\mathrm{VAC}})|$, where $N_{\mathrm{RMP}}$, 
$N_{\mathrm{SFA}}$, and $N_{\mathrm{VAC}}$ are the total number of 
events in the core-crossing bin for the realistic matter 
profile (RMP), the step function approximation (SFA), and vacuum 
oscillation (VAC), respectively. The defined quantity describes
the error of the step function approximation compared to the absolute magnitude
of the matter effects. The error made using the step function 
approximation is at the level of 5\%. Note that the total
matter effects in the electron neutrino-like event rate of such an 
experiment are itself small ($\sim 10\%$). However, future muon 
storage ring neutrino experiments \cite{geer98} promise better 
access to matter effects \cite{nufacmatter}. With such experiments it
will possibly be necessary to take into account details of the matter density 
profile of the Earth.

\vspace*{1pt}\textlineskip	
\section{Summary and Conclusions}	
\label{sec:SC}
\vspace*{-0.5pt}
\noindent
We have studied the quality of a mantle-core-mantle step function
approximation for the realistic Earth matter density profile in
matter enhanced neutrino oscillations. We considered a two neutrino
and a three neutrino framework and found that the results given by the
model function give an excellent approximation to the results
obtained with the realistic profile. The three neutrino analysis we
performed, allowed us to investigate effects on the $\nu_\mu \to
\nu_\tau$ channel, which are important for the description of the
atmospheric neutrino anomaly. We found a drop of the corresponding
oscillation probability in the resonance region.
A nadir angle dependent analysis of the enhancement showed that a sizable
and clear resonance can be expected for nadir angles around
$34^\circ$, where the neutrinos spend maximal traveling length in
the mantle and do not enter the core.
Finally, we tried to quantify the quality of the step function
approximation on the level of rates within a simulation of an atmospheric
neutrino experiment. The obtained error is at the level of 5\%.

\nonumsection{Acknowledgments}
\noindent
We are grateful to M. Lindner and H. Snellman for useful discussions
as well as S.T. Petcov, E.Kh. Akhmedov, and the Referee for
valuable comments and suggestions.
This work was supported by the Max-Planck-Institut f\"ur Physik
(Werner-Heisenberg-Institut) and the Royal Institute of Technology
(KTH) Central Internationalization Funds. Support for this work was
also provided by the ``Sonderforschungsbereich 375 f\"ur
Astro-Teilchenphysik der Deutschen Forschungsgemeinschaft'' and the
Engineer Ernst Johnson Foundation. M.F. wishes to thank the Theoretical
Physics, KTH for their warm hospitality.

\nonumsection{References}

\end{document}